\documentclass{optica-article}

\journal{opticajournal} 

\articletype{Research Article}

\usepackage{etoolbox}
\AtBeginEnvironment{table}{\small}      
\renewcommand{\arraystretch}{1.1}       

\usepackage{booktabs}                   

\usepackage{graphicx}
\usepackage{xcolor}
\usepackage{subcaption}
\usepackage{algorithm}
\usepackage{float}
\usepackage{multirow}
\usepackage{amsmath}
\usepackage{amsfonts}
\DeclareMathOperator*{\argmax}{arg\,max}

\begin{document}

\title{Accelerating 4D Hyperspectral Imaging through Physics-Informed Neural Representation and Adaptive Sampling}

\author{
Chi-Jui Ho,\authormark{1}
Harsh Bhakta,\authormark{2}
Wei Xiong,\authormark{2, 3}
and Nicholas Antipa\authormark{1}
}

\address{
\authormark{1}Dept. of Electrical and Computer Engineering, UC San Diego, La Jolla, CA 92093, USA\\
\authormark{2}Dept. of Chemistry, UC San Diego, La Jolla, CA 92093, USA\\
\authormark{3}Materials Science and Engineering Program, UC San Diego, La Jolla, CA, 92093, USA
}

\email{\authormark{*}\{chh009, hhbhakta, w2xiong, nantipa\}@ucsd.edu} 


\begin{abstract*} 
High-dimensional hyperspectral imaging (HSI) enables the visualization of ultrafast molecular dynamics and complex, heterogeneous spectra. However, applying this capability to resolve spatially varying vibrational couplings in two-dimensional infrared (2DIR) spectroscopy, a type of coherent multidimensional spectroscopy (CMDS), necessitates prohibitively long data acquisition, driven by dense Nyquist sampling requirements and the need for extensive signal accumulation. To address this challenge, we introduce a physics-informed neural representation approach that efficiently reconstructs dense spatially-resolved 2DIR hyperspectral images from sparse experimental measurements. In particular, we used a multilayer perceptron (MLP) to model the relationship between the sub-sampled 4D coordinates and their corresponding spectral intensities, and recover densely sampled 4D spectra from limited observations. The reconstruction results demonstrate that our method, using a fraction of the samples, faithfully recovers both oscillatory and non-oscillatory spectral dynamics in experimental measurement. Moreover, we develop a loss-aware adaptive sampling method to progressively introduce potentially informative samples for iterative data collection while conducting experiments. Experimental results show that the proposed approach achieves high-fidelity spectral recovery using only $1/32$ of the sampling budget, as opposed to exhaustive sampling, effectively reducing total experiment time by up to 32-fold. This framework offers a scalable solution for accelerating any experiments with hypercube data, including multidimensional spectroscopy and hyperspectral imaging, paving the way for rapid chemical imaging of transient biological and material systems.

\end{abstract*}

\section{Introduction}
\label{sec:intro}

Conventional hyperspectral imaging (HSI) captures spectral responses of materials across the spatial domain to encode chemical composition through linear absorption or emission~\cite{goetz1985imaging, chang2003hyperspectral}. Nevertheless, the dimensionality of conventional HSI is insufficient to resolve the transient interactions, couplings, or structural dependencies in complex systems, which often requires additional frequency or temporal dimensions to resolve. Thus, spatially-resolved two-dimensional infrared (2DIR) spectroscopy, a high-dimensional HSI~\cite{zanni2020hyperspectral} modality, has been developed to extend the standard spatial-spectral data structure into the nonlinear regime~\cite{zanni2020hyperspectral,american2021emerging}, recording frequency-frequency correlations to characterize intermolecular couplings and ultrafast dynamics~\cite{ostrander2016spatially}.

However, the acquisition of such high-dimensional data cubes is governed by a fundamental trade-off between resolution, signal-to-noise ratio (SNR), and acquisition speed~\cite{hagen2013review, gehm2007single}.  This bottleneck is particularly acute in 2DIR, where resolving complex dynamics necessitates exhaustive scanning along both the coherence and population time axes. To sample greater than the Nyquist criteria, hundreds of discrete steps are required per dimension, and when integrated with spatial mapping, the data volume grows exponentially. Furthermore, achieving sufficient SNR for these weak nonlinear signals demands long accumulation times at every coordinate to average out noise~\cite{helbing2010compact}. Thus, the combined requirements of signal averaging and high-resolution scanning necessitate hours to capture a single 2DIR data cube ~\cite{ostrander2016spatially}; when extended across multiple waiting times or spatial coordinates, total experiment times frequently reach several days~\cite{bhattacharya2017accelerating}.

To accelerate acquisition, existing methods reconstruct fully sampled data from undersampled measurements via model-based or data-driven frameworks. Model-based methods leverage signal sparsity within compressed sensing frameworks \cite{sanders2012compressed, dunbar2013accelerated, humston2017compressively} or low-rank properties \cite{bhattacharya2017accelerating} to recover full vibrational spectra from sparse measurements. Data-driven approaches, including dynamic mode decomposition \cite{xu2023cutting} and generative adversarial networks \cite{al2022generative}, learn complex signal distributions to denoise spectra. While effective, these strategies are restricted to predefined model dimensions and lack the flexibility to incorporate novel experimental axes. In contrast, our approach supports arbitrarily high dimensionality. Furthermore, data-driven methods are inherently tied to discrete grids and demand massive, fully sampled training datasets, a severe bottleneck given the scarcity of 2DIR data \cite{al2022generative}. By fitting a single measurement, our network bypasses external data requirements and models the signal as a continuous, grid-free function.

\begin{figure*}[t]
     \centering
\includegraphics[width=0.95\textwidth]{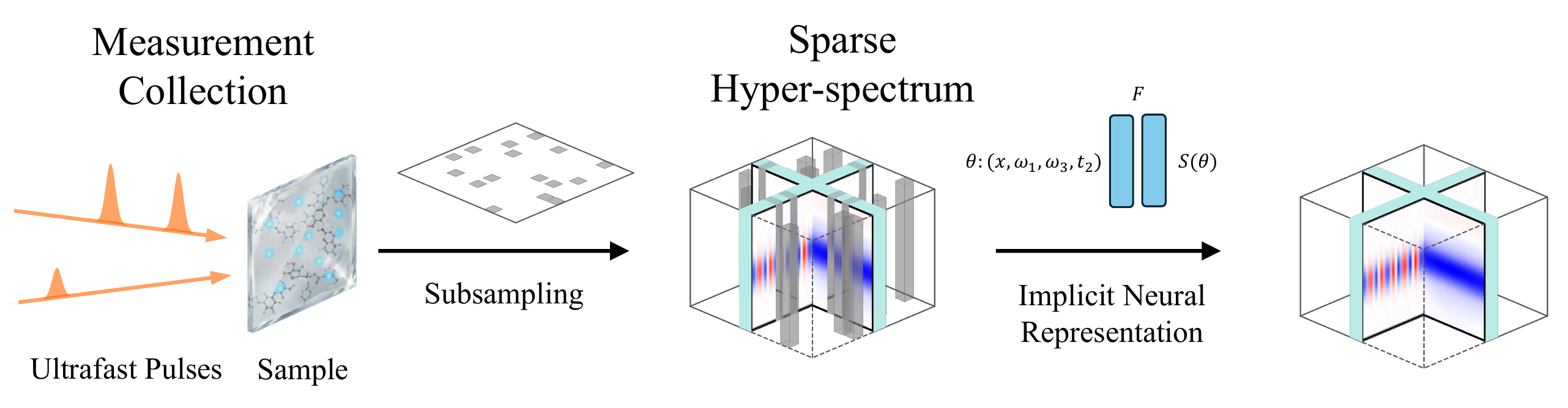}
     \caption{\textbf{Hyper-spectrum reconstruction through neural rendering.} We sparsely sample the hyper-spectrum collected from the interaction between ultrafast pulses and samples, and then use the proposed neural rendering framework to reconstruct the dense hyper-spectrum data in both oscillatory and non-oscillatory regimes.}
     \vspace{-3mm}
     \label{fig:teaser}
\end{figure*}

To address these limitations, we leverage Implicit Neural Representations (INRs) \cite{chen2019learning, park2019deepsdf}, which model hyperspectral data as a continuous field rather than a discrete grid of isolated measurements. As illustrated in Fig.~\ref{fig:teaser}, the INR is implemented via a Multi-Layer Perceptron (MLP). An MLP consists of a network of artificial neurons that maps input coordinates to signal intensities through successive layers of weighted linear sums and nonlinear functions. By optimizing these weights to fit the sampled measurements, the MLP acts as a continuous nonlinear regressor capable of interpolating the signal at unsampled coordinates. To capture rapidly varying temporal dynamics through a physics-informed approach \cite{raissi2019physics, hao2022physics, ho2025differentiable}, we fit the spectrum in the frequency domain and apply an Inverse Real Fourier Transform (IRFT) to compute the corresponding time-domain signal. Furthermore, the INR serves as a nonlinear denoiser, generating high-fidelity spectra from measurements with low signal-averaging counts, reducing the data-collection demands. Complementing this representation, we introduce an adaptive sampling algorithm inspired by dynamic importance sampling \cite{martel2021acorn}. This approach uses the representation error as a surrogate for uncertainty, progressively allocating the sampling budget to regions with the highest mismatch to maximize efficiency.

\noindent Our main contributions are as follows:
\begin{itemize}
    \item We propose a coordinate-based neural field framework for the continuous, data-efficient reconstruction of multidimensional IR spectra from sparse measurements.
    \item Our proposed system achieves high-fidelity spectral recovery despite low signal-averaging counts and undersampling along both coherence and population time axes, yielding up to a 32-fold increase in data efficiency.
    \item We develop a loss-driven adaptive sampling algorithm that prioritizes measurements in regions of high reconstruction sensitivity, enabling targeted and interpretable refinement.
\end{itemize}


\section{Materials and Methods}
\label{sec:mm}

\begin{figure*}[t]
    \centering
    \begin{subfigure}[b]{0.29\textwidth}
        \centering
        \includegraphics[width=\textwidth]{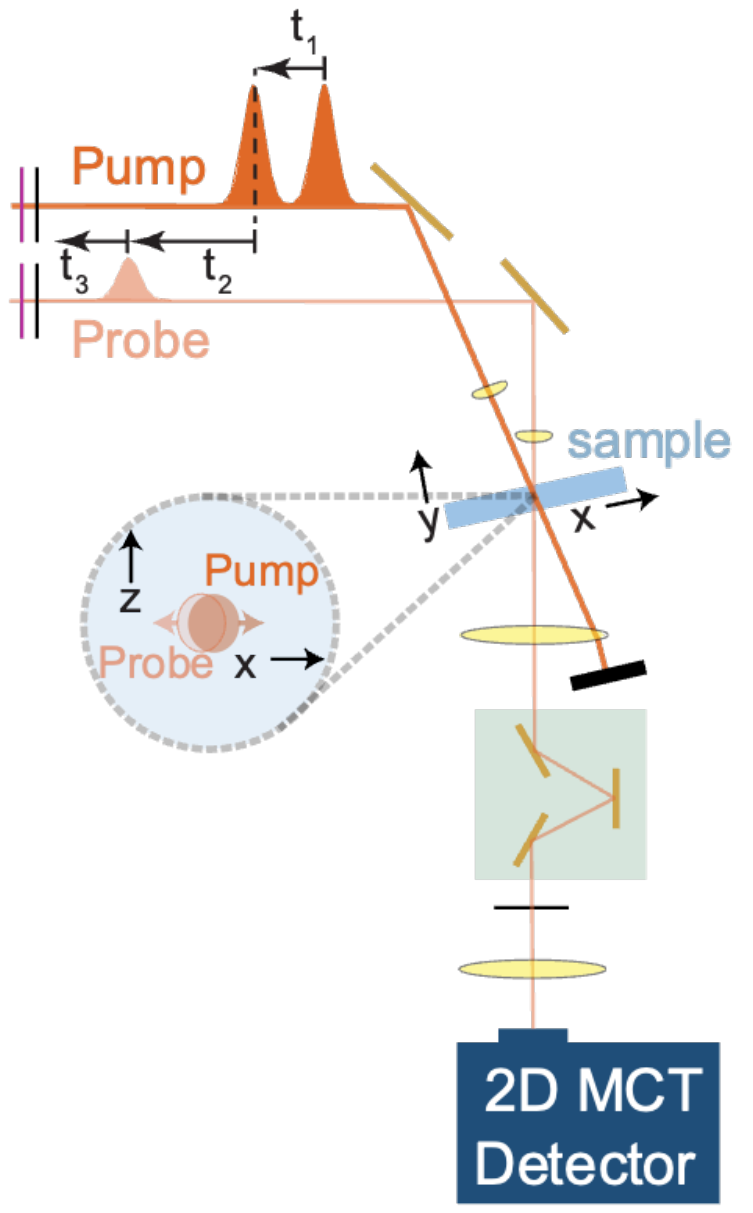}
        \caption{System diagram.}
        \label{fig:dim_a}
    \end{subfigure}
    \hfill 
    \begin{subfigure}[b]{0.68\textwidth}
        \centering
        \includegraphics[width=\textwidth]{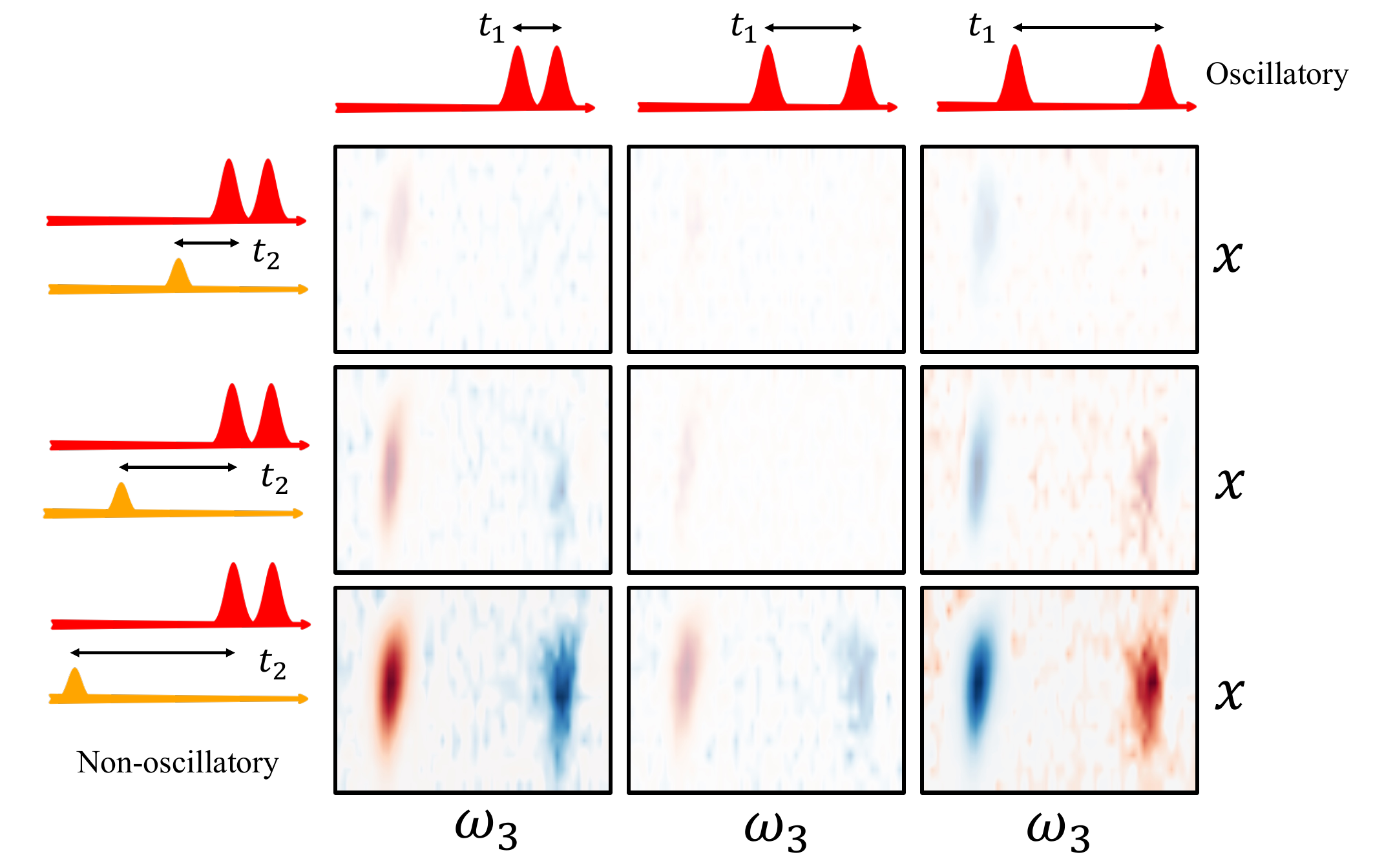}
        \caption{Illustrations of different dimensions.}
        \label{fig:dim_b}
    \end{subfigure}

    \caption{\textbf{Dimensionality of Spatially-Resolved 2DIR.} a) The system diagram. b) Illustration of the four coordinate axes defining the hyperspectral volume. Each 2DIR frame parameterized by $(\omega_1, \omega_3)$, while the temporal and spatial variations are recorded in different waiting time $t_2$ and spatial position $x$.}
    \vspace{-2mm}
    \label{fig:dimension}
\end{figure*}

\subsection{Theoretical Framework and Acquisition of Spatially-Resolved 2DIR}
\label{subsec:mdir_theory}

HSI captures spectral signatures at every spatial coordinate to characterize material composition. While standard HSI efficiently assigns a static 1D linear spectrum to each pixel, the characterization of complex systems often necessitates a nonlinear optical technique to capture underlying molecular interactions and time-dependent evolution. To fulfill this capability, spatially-resolved 2DIR serves as a multidimensional spectroscopic expansion of the HSI framework \cite{zanni2020hyperspectral}. Similar to regular 2DIR \cite{hamm1998structure,mukamel1995principles}, this advanced hyperspectral modality probes the nonlinear vibrational manifold to resolve the structural dynamics and molecular couplings within the system.

\subsubsection{Nonlinear Signal Generation and Temporal Evolution}

To disentangle the specific vibrational couplings and relaxation processes that constitute these dynamics, 2DIR measures the nonlinear response of a material using a pulse sequence consisting of three ultrafast infrared pulses in a pump-probe geometry with distinct time intervals, as shown in Fig. ~\ref{fig:dim_a}:
\begin{enumerate}
    \item \textbf{Coherence Time ($t_1$):} The delay between the first two excitation (pump) pulses, modulated via a pulse shaper, during which the system evolves in vibrational coherence states.
    \item \textbf{Population Time ($t_2$):} The waiting time between the second pump pulse and a weaker probe pulse, controlled by a motorized mechanical delay stage, to capture non-oscillatory dynamics such as vibrational relaxation, and sometimes, oscillatory dynamics if coherence states are involved.
    \item \textbf{Detection Time ($t_3$):} The temporal axis of the emitted signal following the probe pulse, describing the free-induction decay of the nonlinear response.
\end{enumerate}
\noindent During these intervals, the evolution of coherent and population states is recorded through the nonlinear optical signal; this signal is subsequently dispersed by a grating spectrometer, which encodes the spectral information as a function of probe frequency ($\omega_3$). To visualize the correlation between states excited before and after $t_2$, a Fourier Transform is performed along the coherence time $t_1$ to yield the pump axis $\omega_1$. This transformation results in a pump-probe correlation spectrum $S(\omega_1, \omega_3)$.

\subsubsection{Spatial Resolution and Acquisition Dimensions}

In addition to spectral dimensions, spatially-resolved 2DIR incorporates a spatial dimension ($x$) via a monochromator slit, which constrains the input field to select a line of an image of the sample \cite{yang2023enabling}. This configuration enables the measurement to resolve the spatiotemporal evolution of molecular states. Consequently, the spectroscopic measurement is partitioned into four dimensions $(\omega_1, \omega_3, t_2, x)$, encoding molecular couplings, energy exchange, and relaxation dynamics, and is ultimately collected by a HgCdTe Focal Plane Array (PhaseTech). 

Our optimization efforts focus specifically on the controllable acquisition dimensions $(t_1, t_2)$ rather than the natively resolved axes. As shown in Fig. \ref{fig:dimension}, the detector configuration naturally disperses and multiplexes the spatial and detection frequency dimensions $(x, \omega_3)$ onto the array, whereas the temporal dimensions $(t_1, t_2)$ have to be exhaustively scanned.


\subsubsection{Acquisition Protocol.} 
To collect the full 2D IR hyperspectral data, we scan the coherence time $t_1$ from 0 to 8000~fs in 32~fs steps, and the population time $t_2$ from -6~ps to 12~ps in 0.2~ps steps. To overcome bandwidth limitations imposed by the pulse shaper resolution, we implement a rotating frame strategy, which down-shifts the oscillation frequency of the measured coherence signal and permits accurate sampling at larger time steps. The probe frequency $\omega_3$ spans 1906.6--2058.7~cm$^{-1}$ with a resolution of $1.197 \pm 0.001$~cm$^{-1}$/pixel. The spatial axis ($x$) covers a field of view of at least $\pm 200$~$\mu$m with a resolution of $24 \pm 6$~$\mu$m/pixel, depending on the system configuration. Finally, to enhance the Signal-to-Noise Ratio, measurements at each $(t_1, t_2)$ coordinate are averaged over 20 accumulations ($r$). On average, it takes 2 hours and 10 minutes to collect a four-dimensional array with coordinates $(t_2, x, \omega_1, \omega_3)$ using $r=20$. In this study, we collect a total of five spatially-resolved 2DIR datasets, acquired under different rotation angles, which determines the quality and noise level of the collected spectra \cite{xiang2022ultrafast, xiang2018two, xiang2019manipulating}.

\subsubsection{Generalization to Coherent Multidimensional Spectroscopy.}
The resulting 4D dataset poses a reconstruction challenge inherent to 2DIR spectroscopy, where signal dimensions exhibit two distinct behaviors: oscillatory evolution arising from coherent states ($t_1$), and non-oscillatory decay or variation associated with population dynamics and spatial distribution ($t_2, x$). Because our reconstruction approach addresses these fundamental signal dynamics, it offers a generalized methodology applicable to various CMDS techniques, such as bioimaging \cite{dicke2021application, tiwari2018spatially} and transient absorption microscopy \cite{zhu2020transient, gross2023progress}.

\subsection{Neural Representation Algorithm}
\label{subsec:algo}

To model multidimensional IR spectroscopic dynamics in unsampled regions, we use INRs to represent the data as a continuous function of its acquisition coordinates, rather than a discrete array of data points. By learning the underlying mapping between these experimental coordinates and the measured signal intensities, INRs are able to estimate continuous signal values even under low subsampling rates of the population ($t_2$) and coherence ($t_1$) time axes.

\subsubsection{Implicit Neural Representation}

INRs offer an alternative to conventional methods for modeling high-dimensional data \cite{park2019deepsdf, chen2019learning}. Traditionally, signals are explicitly represented as discrete arrays of pixels or voxels, with values stored on a predefined resolution grid. In contrast, INRs bypass this discrete grid by formulating a signal as a continuous function of its coordinates. With INRs, signals are no longer constrained by a fixed sampling rate but can be evaluated at any arbitrary coordinate, making the representation inherently resolution-independent. To parameterize this continuous function, INRs typically employ an MLP, which naturally maps continuous domains to complex outputs without requiring rigid, predefined analytical equations. Specifically, the MLP takes coordinates as inputs and generates the corresponding signal value at that exact location. The relationship between coordinates and signal values is parameterized by the tunable weights and biases across sequential fully-connected layers of the network. Through a series of matrix multiplications and nonlinear activations, the MLP predicts the signal values at the input coordinates. Unlike supervised data-driven approaches \cite{he2016deep} that rely on extensive pre-training, the MLP is optimized from scratch for each specific experimental instance. This instance-specific optimization allows the MLP to learn the continuous topology of a single signal without the need for large external datasets, ensuring high-fidelity reconstruction from limited samples.



\subsubsection{Network Architecture and Training Strategy}
\label{subsec:mlp}
\begin{figure}[H]
     \centering
     \includegraphics[width=1.0\textwidth]{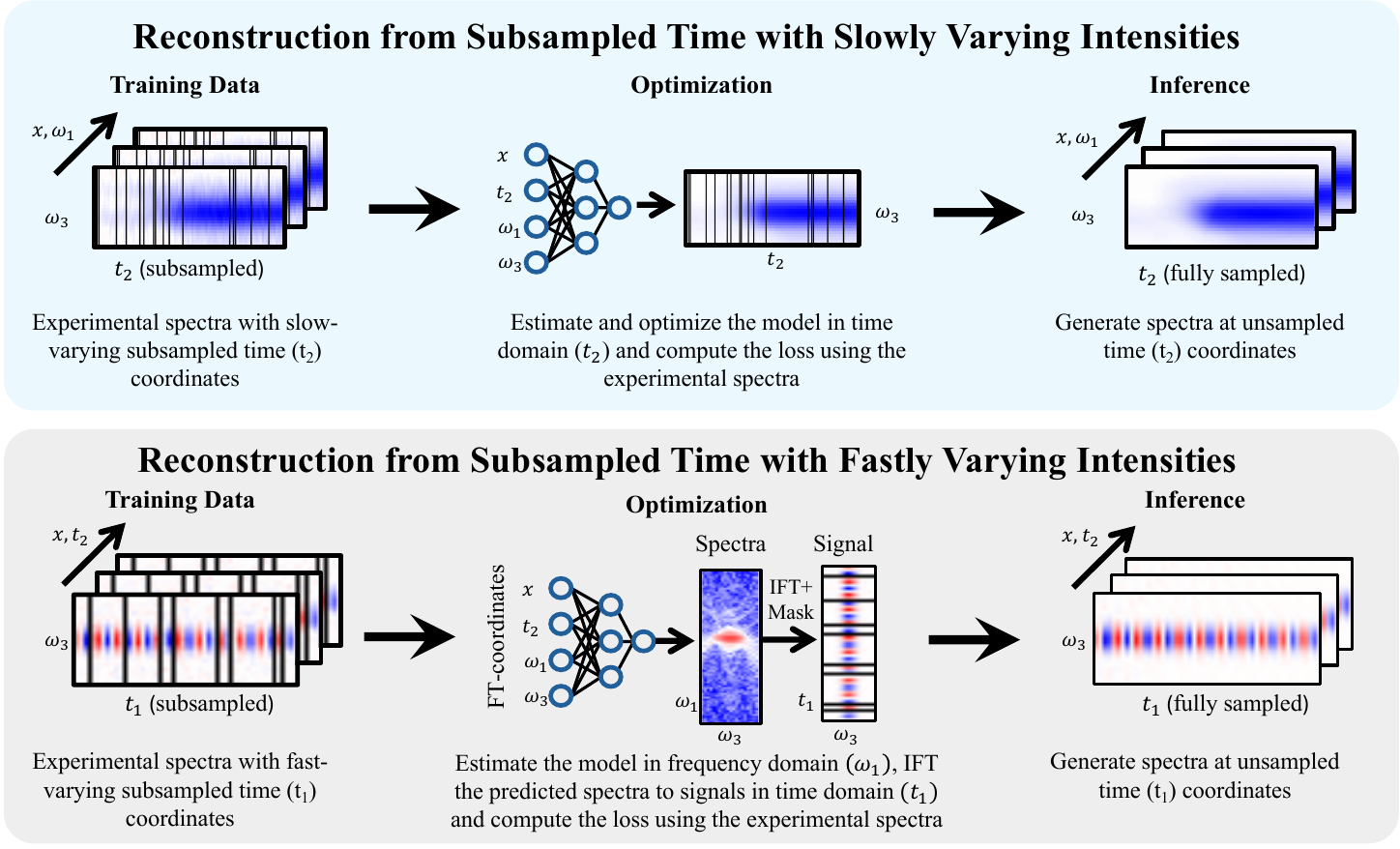}
     \caption{\textbf{Our MLP-based neural representation model.} We construct neural representations for signals subsampled along either the population (slow varying) or coherence (fast varying) time dimensions. \textbf{Population:} A coordinate-based MLP is trained to learn the temporal dynamics along population time from sparsely sampled data and to generalize this knowledge to unsampled points during inference. \textbf{Coherence:} The MLP is optimized in the frequency domain, after which an IRFT is applied to recover estimated raw signals for loss computation, excluding intensities at unsampled coordinates. During inference, the trained network reconstructs rapidly varying 4D signals across the full coordinate domain.}
     \vspace{-3mm}
     \label{fig:MLP}
\end{figure}

As visualized in Fig.~\ref{fig:MLP}, we employ a coordinate-based MLP to map normalized spatial ($x$), temporal ($t_1, t_2$), and frequency ($\omega_1, \omega_3$) coordinates to their corresponding spectroscopic intensities. The network is designed to accurately model both rapidly oscillating and slowly varying signal components. Specifically, the MLP consists of four hidden layers, each comprising 64 neurons with Rectified Linear Unit (ReLU)~\cite{agarap2018deep} activations, followed by a final linear output layer that yields the predicted frequency-domain signal intensity $\hat{S}$.

\noindent \textbf{Slowly Varying Intensities.}
Given that temporal dynamics along the population time axis ($t_2$) are inherently smooth, the MLP directly models the relationship between the input coordinates and the associated intensities. In this regime, the intensity reconstruction is driven by computing the Mean Squared Error (MSE) directly between the MLP prediction $\hat{S}$ and the ground truth frequency-domain measurements $S$ across the set of sampled coordinates $\Theta$:
\begin{equation}
    \mathcal{L}_{\mathrm{MSE, slow}} = \frac{1}{|\Theta|} \sum_{(x, t_2, \omega_1, \omega_3) \in \Theta} \Big( \hat{S}(x, t_2, \omega_1, \omega_3) - S(x, t_2, \omega_1, \omega_3) \Big)^2,
\end{equation}
where $|\Theta|$ denotes the total number of sampled points. At the inference stage, the trained MLP takes densely sampled 4D coordinates as input to predict intensities at both sampled and unsampled locations.

\noindent \textbf{Rapidly Varying Intensities.}
Conversely, standard MLPs struggle to model signals exhibiting rapid oscillations, a characteristic natural to the coherence time axis ($t_1$); this difficulty is exacerbated under sparse sampling conditions. To address this, we adopt a physics-informed strategy. Rather than directly fitting the MLP to rapidly oscillating time-domain signals, the MLP is queried to output the frequency-domain spectrum $\hat{S}(x, t_2, \omega_1, \omega_3)$. We then apply an Inverse Real Fourier Transform (IRFT) along the $\omega_1$ axis to recover the corresponding time-domain signal estimate, denoted as $\hat{s}(x, t_2, t_1, \omega_3) = \mathcal{F}^{-1}_{\mathrm{R}}\{ \hat{S} \}$. The loss is subsequently computed in the time domain by comparing these estimates against the sparse time-domain measurements $s$:
\begin{equation}
    \mathcal{L}_{\mathrm{MSE, fast}} = \frac{1}{|\Theta|} \sum_{(x, t_2, t_1, \omega_3) \in \Theta} \Big( \hat{s}(x, t_2, t_1, \omega_3) - s(x, t_2, t_1, \omega_3) \Big)^2.
\end{equation}
Leveraging the inherent low-frequency bias of MLPs as a regularizer \cite{rahaman2019spectral}, the model learns the smoother spectral function $\hat{S}$ to preserve time-domain oscillations while resisting noise and sparse-acquisition artifacts.

\begin{figure}[t]
     \centering
     \includegraphics[width=0.6\textwidth]{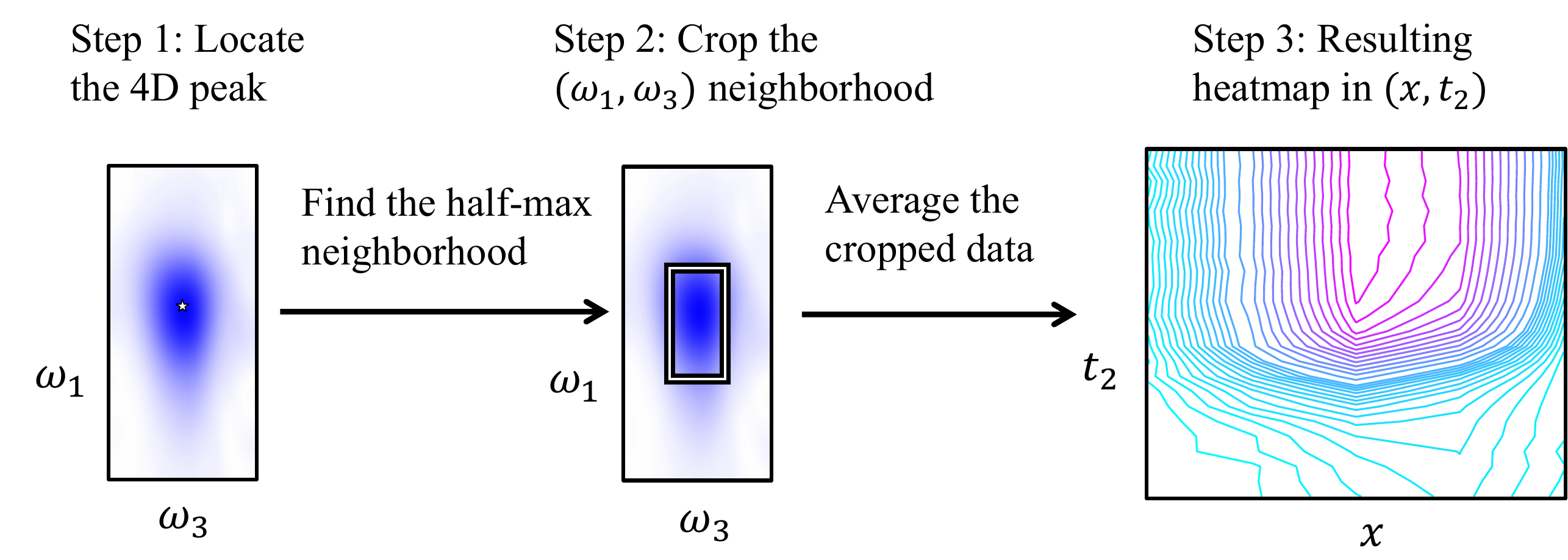}
     \caption{\textbf{2D representation of 4DIR data.} 
     A bounding box is defined around the dominant peak in each $(\omega_1,\omega_3)$ slice. The intensities enclosed within this region are averaged to reduce the four-dimensional data to a two-dimensional representation along the $(x,t_2)$ axes, which is subsequently used for evaluation.}
     \vspace{-4mm}
     \label{fig:projection}
\end{figure}

\paragraph{Statistical-Moment-Matching Constraints.}
To capture spatial dynamics accurately along with the spectral intensity \cite{akselrod2014subdiffusive, penwell2017resolving, pandya2022tuning, balasubrahmaniyam2023enhanced, sandik2025cavity, xu2023ultrafast}, we apply constraints on the statistical moments ($\mu, \sigma$) of the spatial variable ($x$) as follows. As shown in Fig.~\ref{fig:projection}, we first reduce the 4D spectra $S(x, t_2, \omega_1, \omega_3)$ to a 2D spatial-temporal profile $M(x, t_2)$ with the following three steps: a) identify the peak of intensities, b) isolate the half-maximum neighborhood around this peak in the $(\omega_1, \omega_3)$ frequency plane, and c) average the intensities within this region. After this dimension reduction, we normalize the spatial profile $M(\cdot, t_2)$ for each sampled population time $t_2$ to treat it as a probability distribution:
\begin{equation}
    p_M(i \mid t_2) = \frac{M(x_i, t_2)}{\sum_{j=1}^{N_x} M(x_j, t_2)}, \quad i=1, \dots, N_x,
\end{equation}
where $x_i$ represents the spatial coordinate at index $i$, and $N_x$ is the total number of spatial coordinates. This formulation allows us to analyze the spatial evolution via the mean position $\mu_M$ and standard deviation $\sigma_M$:
\begin{equation}
\label{eq:mean_std}
    \mu_M(t_2) = \sum_{i=1}^{N_x} x_i \, p_M(i \mid t_2), 
    \quad 
    \sigma_M(t_2) = \sqrt{\sum_{i=1}^{N_x} x_i^2 \, p_M(i \mid t_2) - \mu_M(t_2)^2}.
\end{equation}
Thus, the moment-matching loss ensures that the predicted profile $\tilde{M}$ matches the spatial moments of the reference $M$:
\begin{equation}
    \mathcal{L}_{\mathrm{moment}} = \sum_{t_2 \in \mathcal{T}_2} \Big( 
    |\mu_{\tilde{M}}(t_2) - \mu_M(t_2)|^2 
    + |\sigma_{\tilde{M}}(t_2) - \sigma_M(t_2)|^2 
    + \|\tilde{M}(\cdot, t_2) - M(\cdot, t_2)\|_2^2 
    \Big).
\end{equation}

\paragraph{Regularization.}
To pursue physically accurate dynamics, we apply regularization functions to the spatial variance to encourage smooth and monotonic changes. These functions act as penalty terms that constrain the solution space, favoring models that align with prior structural assumptions \cite{willoughby1979solutions, aster2018parameter}. Let $\mathcal{T}_2$ represent the set of all sampled time points along the population time axis ($t_2$), containing a total of $K = |\mathcal{T}_2|$ sampled points. We index these sampled time points in strictly chronological order, denoting the $k$-th time point as $t_{2_k}$ and apply a hinge-type monotonicity constraint by:

\begin{equation}
\label{eq:mono}
\mathcal{L}_{\mathrm{mono}} = \frac{1}{K-1} \sum_{k=1}^{K-1} \max\big\{0, \, \delta - [\sigma_M(t_{2,k+1}) - \sigma_M(t_{2,k})] \big\},
\end{equation}

\noindent where $\delta \ge 0$ is a margin parameter. Additionally, to prevent numerical instability and ensure smooth temporal evolution, we penalize positive second-order finite differences:

\begin{equation}
\mathcal{L}_{\mathrm{smooth}} = \frac{1}{K-2} \sum_{k=1}^{K-2} \max\big\{ 0, \, \sigma_M(t_{2,k+2}) - 2\sigma_M(t_{2,k+1}) + \sigma_M(t_{2,k}) - \gamma \big\},
\end{equation}

\noindent where $\gamma$ serves as a threshold for curvature.

\paragraph{Total Objective.}
The final loss function is a weighted sum of all these components:
\begin{equation}
    \mathcal{L} = \lambda_{\mathrm{MSE}}\mathcal{L}_{\mathrm{MSE}} + \lambda_{\mathrm{moment}}\mathcal{L}_{\mathrm{moment}} + \lambda_{\mathrm{mono}}\mathcal{L}_{\mathrm{mono}} + \lambda_{\mathrm{smooth}}\mathcal{L}_{\mathrm{smooth}},
\end{equation}

where the coefficients $\lambda$ are empirical hyperparameters, balancing the contribution of each term during optimization.

\subsection{Data Preparation}
\label{subsec:data_prep}

Using the 4D hypercube dataset described in Sec.~\ref{subsec:mdir_theory}, we simulate and pre-process measurements acquired under reduced sampling budgets as follows.

\noindent \textbf{Signal accumulation ($r$).} As mentioned above, each experimental spectrum is produced by averaging 20 repeated measurements. To simulate data obtained in different data aggregation counts $r$, we average the spectrum with a subset of repeatedly sampled data.

\noindent \textbf{Temporal subsampling.} Along the temporal axes $t_1$ and $t_2$, we randomly or periodically subsample the measurements to emulate data acquisition with lower temporal sampling rates. The remaining axes are kept fully sampled.

\noindent \textbf{Normalizations.}
In 2DIR spectra, the positive and negative signals may differ in amplitude by up to an order of magnitude. This imbalance can make it hard to recover any weaker peaks. To address this, we rescale the weaker peak to match the dynamic range of the stronger one prior to model fitting, and scale it back at the inference stage. In model training and inference, all input coordinates are normalized to $[0,1]$ to maintain consistent scaling across all dimensions.

\subsection{Adaptive Sampling}
\label{subsec:adaptive_method}

\begin{figure}[t]
     \centering
     \includegraphics[width=0.9\textwidth]{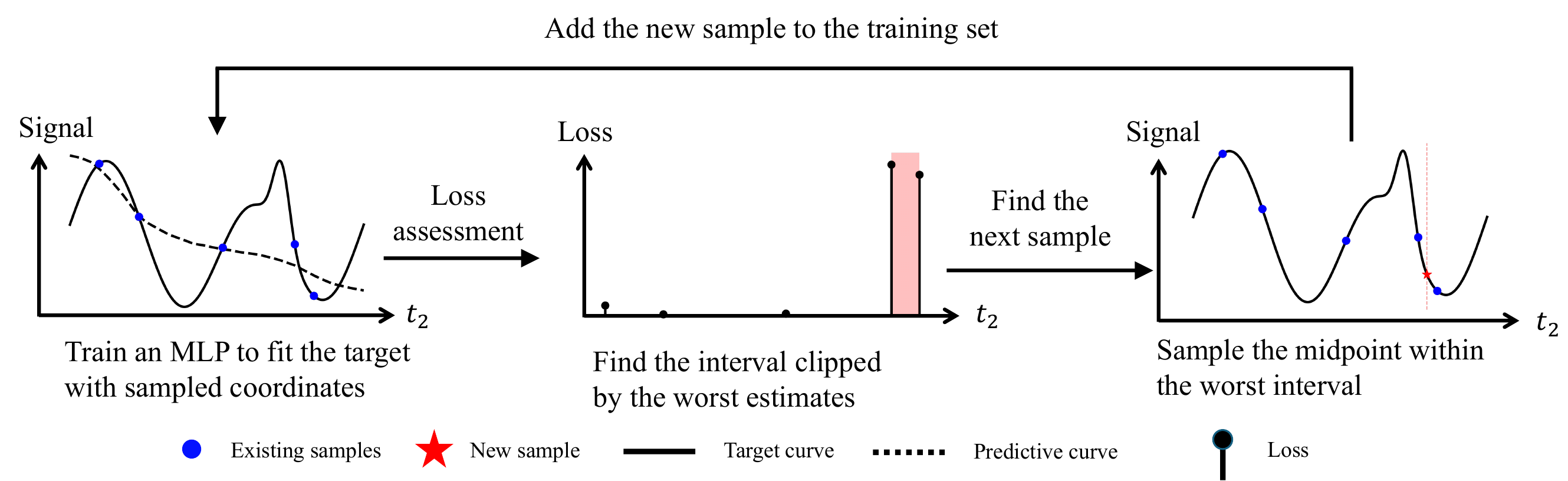}
     \caption{\textbf{Our loss-driven adaptive sampling method.} Beginning with the initially sampled coordinates (blue dots), the MLP is trained to model the spectral dynamics and the loss is evaluated along the population time $t_2$. The interval exhibiting the highest average prediction error is identified, and its midpoint is selected as the next sample.}
     \vspace{-2mm}
     \label{fig:adaptive}
\end{figure}

We apply a loss-aware adaptive sampling strategy~\cite{martel2021acorn} to improve data efficiency along the population time ($t_2$) axis. As illustrated in Fig.~\ref{fig:adaptive}, the procedure first acquires measurements at a sparse set of predefined $t_2$ locations to establish an initial set of samples. To initiate the adaptive sampling phase, we fit the MLP to these initial spectra and compute the reconstruction loss against the experimental measurements. To quantify uncertainty at specific time delays, this loss is aggregated across the spatial and frequency dimensions, yielding an error metric $\mathcal{L}(t_{2,k})$ for each sampled point, where $t_{2,k}$ denotes the $k$-th sample in the chronologically ordered sequence.

Using this metric, we compute the average reconstruction error, $s_k$, for the interval between each pair of adjacent samples, $t_{2,k}$ and $t_{2,k+1}$:
\begin{equation}
    s_k = \frac{1}{2}\big(\mathcal{L}(t_{2,k}) + \mathcal{L}(t_{2,k+1})\big), \quad k=1,\ldots,K-1{.}
\label{eq:score-mean}
\end{equation}

The averaged error $s_k$ serves as a heuristic for uncertainty within an interval. Large estimation errors at the boundaries suggest that the underlying dynamics of that region are poorly captured. To maximize the information gain of the next measurement, we identify the interval with the highest error, $k^\star = \arg\max_k s_k$, and select its midpoint as the subsequent sampling target.

\begin{equation}
    \label{eq:midpoint}
    t_2^\star = \frac{1}{2}\big(t_{2,k^\star} + t_{2,k^\star+1}\big){,}
\end{equation}

\noindent which is rounded to the nearest available delay. The MLP is subsequently refined using this newly acquired measurement, and sampling and refinement are alternated until the predefined sampling budget is exhausted. By iteratively bisecting the intervals bounded by the highest reconstruction errors, the algorithm progressively closes interpolation gaps. This yields a dynamic sampling distribution naturally optimized to capture the underlying physical dynamics.

\subsection{Software Implementation Details}

We use PyTorch as our optimization framework~\cite{paszke2019pytorch} and drive the network optimization by the Adam algorithm \cite{kingma2014adam} with parameters $(\beta_1, \beta_2) = (0.9, 0.999)$. Data representation performance is evaluated both qualitatively and quantitatively. For qualitative evaluation, we visualize the $(\omega_3,\omega_1)$ and $(\omega_3, t_2)$ slices around the peak of 4D data. Quantitatively, we report the overall MSE and three population time profiles: the peak-intensity, mean, and standard-deviation profiles. In particular, the peak intensity at each population time is computed by averaging the magnitudes within the bounding box depicted in Fig. ~\ref{fig:projection}.

\section{Results and Discussion}
\label{sec:rd}

In this section, we first evaluate the networks trained with different numbers of data averaging counts and varying $t_2$ sampling densities. We also report each network's ability to capture rapid variations along the $t_1$ axis and show how the adaptive sampling strategy selects the informative points from unseen candidates. Using the MLP-based representation, we model the spectra with fewer repeated samples (Sec.~\ref{ssec:repeat}), subsampled $t_2$ indices (Sec.~\ref{ssec: subt2}), and $t_1$ indices (Sec.~\ref{ssec: subt1}). We then present the results from different sampling strategies in Sec.~\ref{ssec: adaptive}. 

\begin{figure}[H]
     \centering
     \includegraphics[width=0.9\textwidth]{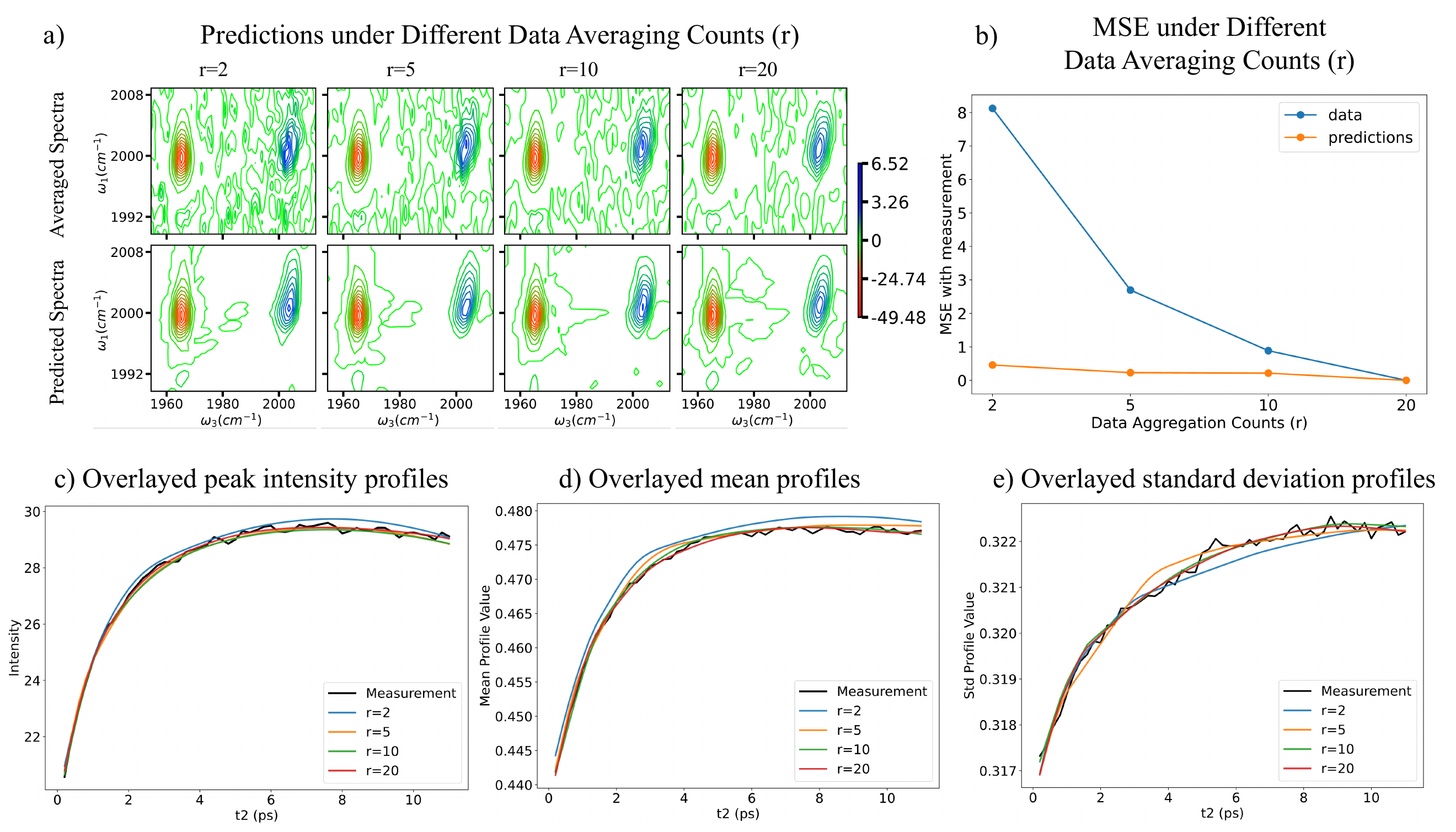}
     \caption{\textbf{Summary of results under different numbers of repeated samplings.} 
     (a) Measurements averaged over varying data averaging counts ($r$) and reconstructed spectra under the same count.
     (c) Using the least noisy case (highest $r$) as reference, we compute the MSE of other measurements and predictions against it.
     (d) Mean and (e) standard deviation profiles of reconstructions at different repeat counts, with the reference shown as a solid black line.
      }
     \label{fig:repeat}
     \vspace{-4mm}
\end{figure}

\begin{table}[H]
\centering
\caption{Quantitative evaluation under different data averaging counts ($r$).}
\label{tab:repeat}
\begin{tabular}{lrrr}
\toprule
$r$ & Intensity MSE & Mean profile MSE & Std profile MSE \\
\midrule
2  & 1.320e-02 & 1.280e-06 & 3.820e-07 \\
5  & 8.330e-03 & 6.600e-07 & 1.240e-07 \\
10 & 8.250e-03 & 6.050e-07 & 9.990e-08 \\
20 & \textbf{4.990e-03} & \textbf{4.360e-07} & \textbf{4.860e-08} \\
\bottomrule
\end{tabular}
\vspace{-3mm}
\end{table}

\subsection{Prediction Stability}

\subsubsection{Low-Accumulation Regimes}
\label{ssec:repeat}

We evaluate the noise-robustness of our framework by varying the signal-averaging counts ($r \in \{2, 5, 10, 20\}$) of input spectra to simulate different noise levels. The spectra with different noise levels serve as the reference measurement to train the MLP. The results are summarized in Fig.~\ref{fig:repeat} and Table~\ref{tab:repeat}. As illustrated in Fig.~\ref{fig:repeat}(a), reducing the signal-averaging count increases the noise level and distorts weaker peaks in the synthesized spectra. In contrast, the MLP is robust to noise and predicts spectra retaining structural integrity, effectively suppressing elevated noise. In Fig.~\ref{fig:repeat}(b), we assess reconstruction performance against the measurement averaged across all available acquisitions, representing the lowest achievable noise level. As we can see, the raw measurements deviate significantly from the reference as the data aggregation count decreases,  while the model predictions remain consistent across different noise levels.

As observed in Table~\ref{tab:repeat} and Fig.~\ref{fig:repeat}(c)–(e), the shape of all three profiles remain stable under different data averaging counts, with moderate mismatch in lower counts. These findings demonstrate that our model is robust to noise, which alleviates the need for extensive data aggregation counts and improves data acquisition efficiency. Notably, the choice of acceptable $r$ is analysis-dependent. In terms of intensity recovery, we can reduce the value to $r=5$ with consistent recovery. Nonetheless, if one pursues accurate mean and standard deviation of the spatial profile, a higher number $r=10$ is required.

\subsubsection{Undersampled Population Temporal Data}
\label{ssec: subt2}

\begin{figure}[t]
     \centering
     \includegraphics[width=\textwidth]{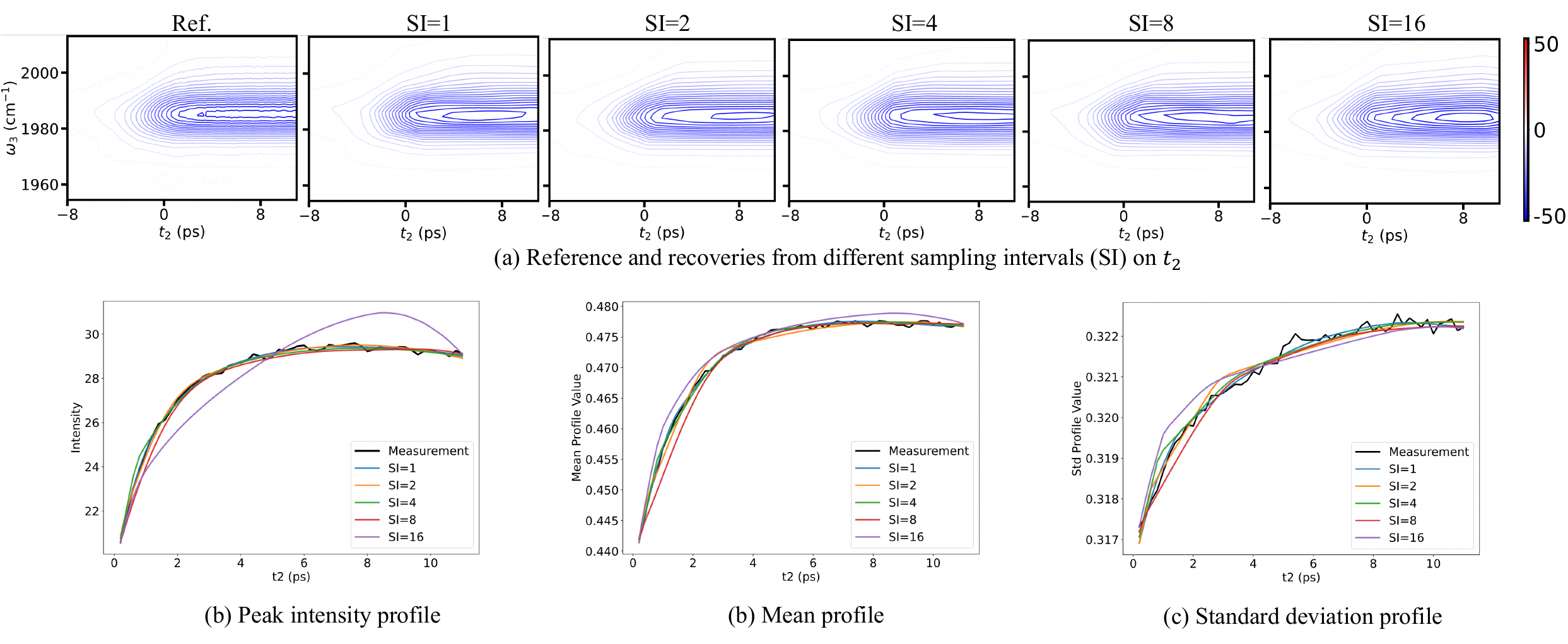}
     \caption{\textbf{Summary of results under different sampling intervals (SI) along the population time.} (a) Reference and reconstructed spectra obtained at various sampling rates along the population time axis.
     (b–d) Peak intensity, mean and standard deviation profiles of reconstructions at different sampling intervals, with the reference shown as a solid black line.}
     \vspace{-4mm}
     \label{fig:t2_results}
\end{figure}

\begin{table}[t]
\centering
\caption{Quantitative evaluation under different sampling intervals (SI) along the population time.}
\label{tab:t2_results}
\begin{tabular}{lrrr}
\toprule
SI & Intensity MSE & Mean profile MSE & Std profile MSE \\
\midrule
1  & \textbf{4.990e-03} & \textbf{4.360e-07} & \textbf{4.860e-08} \\
2  & 5.590e-03 & 5.120e-07 & 6.390e-08 \\
4  & 1.390e-02 & 1.290e-06 & 1.480e-07 \\
8  & 1.680e-02 & 2.860e-06 & 1.570e-07 \\
16 & 1.570e-01 & 3.490e-06 & 3.000e-07 \\
\bottomrule
\end{tabular}
\vspace{-3mm}
\end{table}

We simulated 2DIR data under varying $t_2$ sampling intervals (SIs) via uniform undersampling. With a baseline rate of 0.2 ps/step, SIs ranging from 2 to 16 were evaluated to optimize the MLP. Results across these sampling densities are presented in Fig.~\ref{fig:t2_results} and Table~\ref{tab:t2_results}. As shown in Fig.~\ref{fig:t2_results}(a), the reconstructed spectra faithfully reproduce the spectral structures in unsampled $t_2$ regions regardless of different sampling rates except for $SI=16$. As observed in Fig.\ref{fig:t2_results}(b)-(d) and Table~\ref{tab:t2_results}, the most significant mismatch occurs when the sampling interval is 16, where the sparse sampling rate makes capturing the temporal dynamics extremely difficult. We observe the same behavior in the mean and standard deviation of the spatial profile, which closely match the reference until the sampling interval reaches 16. These results demonstrate the generalization capability of the MLP in unobserved temporal regions, enabling lower sampling rates without compromising reconstruction fidelity. As discussed in Sec.~\ref{ssec:repeat}, the choice of acceptable sampling interval is evaluation dependent.


\begin{figure}[t]
    \centering
    \includegraphics[width=\textwidth]{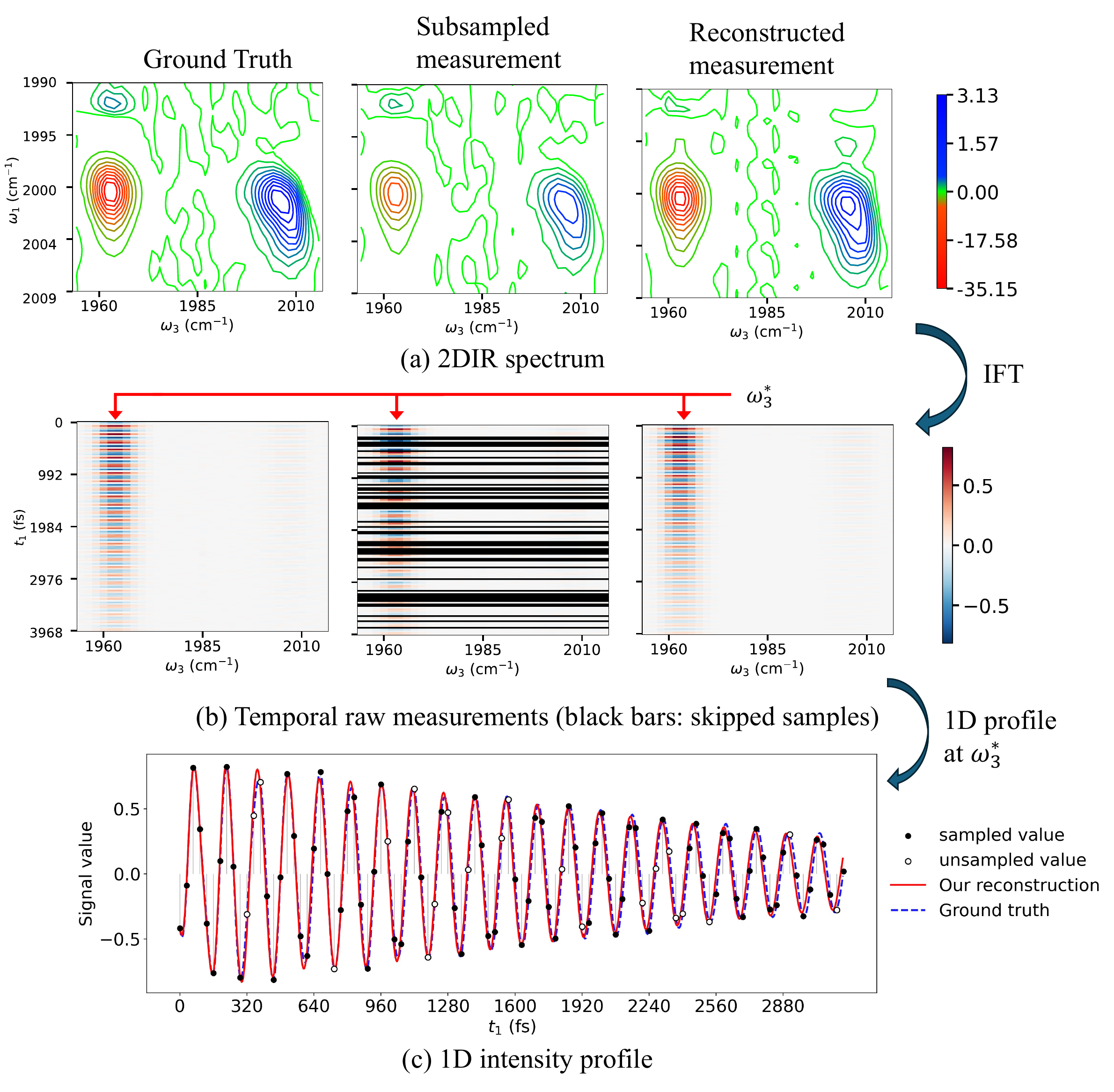} 
    \caption{\textbf{Reconstruction results from subsampled $t_1$ domain data.} 
    (a) Reconstructed 2DIR spectrum compared against the ground truth. (b) Temporal raw measurements showing the fully sampled ground truth and the subsampled input. (c) A 1D intensity profile extracted from (b) demonstrating signal recovery. To visualize oscillations beyond the native samples, the ground truth is densified via sinc interpolation, while the dense prediction is obtained by evaluating the IRFT with a finer $t_1$ sampling.}
    \label{fig:t1_results}
    \vspace{-3mm}
\end{figure}

\begin{figure}[t]
    \centering
    \includegraphics[width=\textwidth]{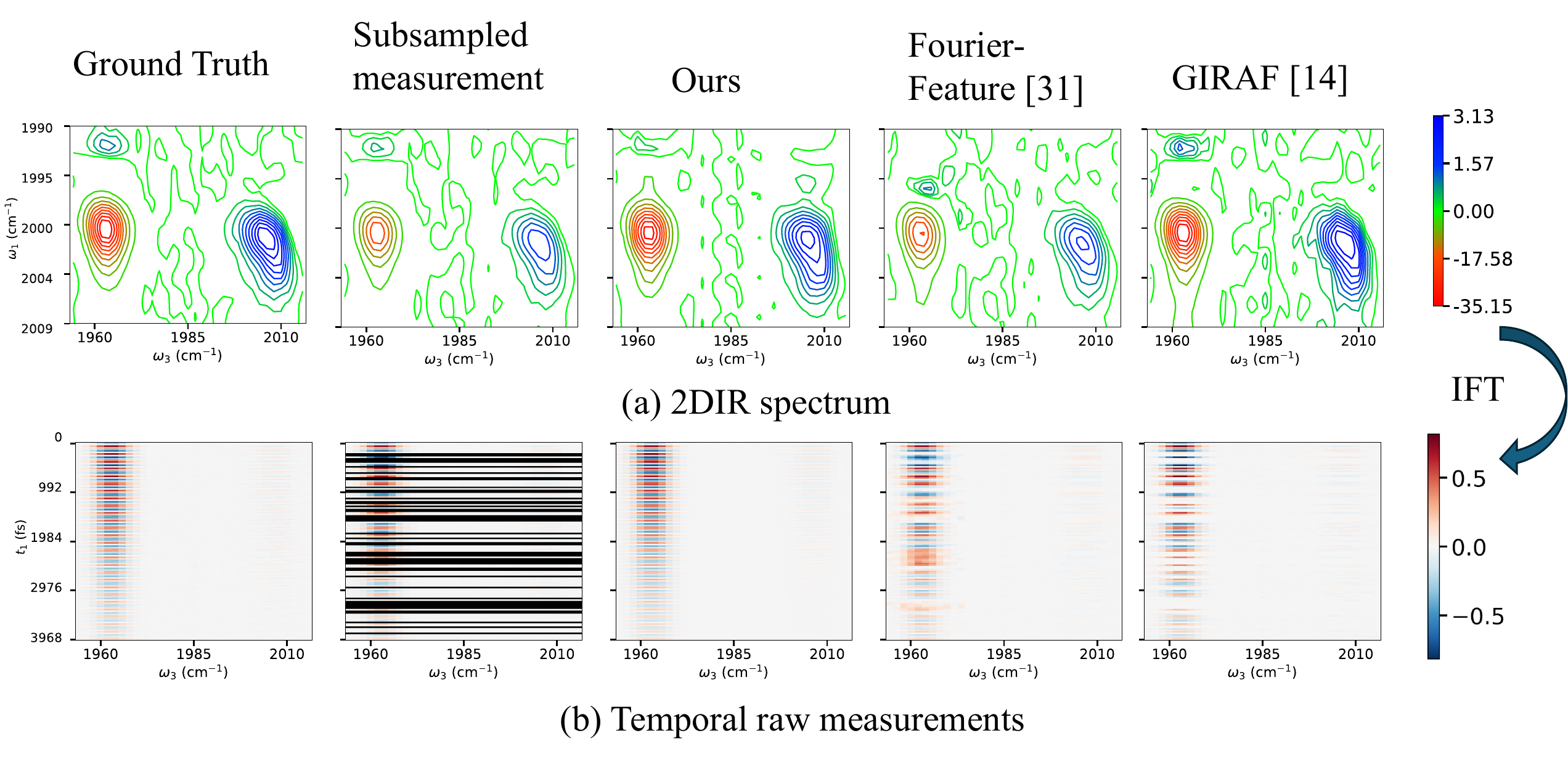}
    \caption{\textbf{Reconstruction results from subsampled $t_1$ domain data using different algorithms.} 
    (a) The 2DIR spectrum and (b) temporal  measurements generated from different reconstruction algorithms compared against the ground truth.}
    \label{fig:t1_cmp}
    \vspace{-3mm}
\end{figure}

\begin{table}[t]
\centering
\caption{Reconstruction performance on undersampled $t_1$ data across different algorithms.}
\label{tab:drop50_avg_ang2_to_ang02}
\begin{tabular}{lrr}
\toprule
Algorithm & Spectrum MSE & Unsampled Temporal MSE\\
\midrule
RFFT & \textbf{1.333e-02} & \textbf{1.583e-04} \\
GIRAF-2D \cite{bhattacharya2017accelerating} & 8.607e-02 & 3.443e-03 \\
Fourier-features~\cite{tancik2020fourier} & 1.615e-01 & 6.190e-03 \\
\bottomrule
\end{tabular}
\end{table}

\begin{figure}[t]
    \centering
    \includegraphics[width=\textwidth]{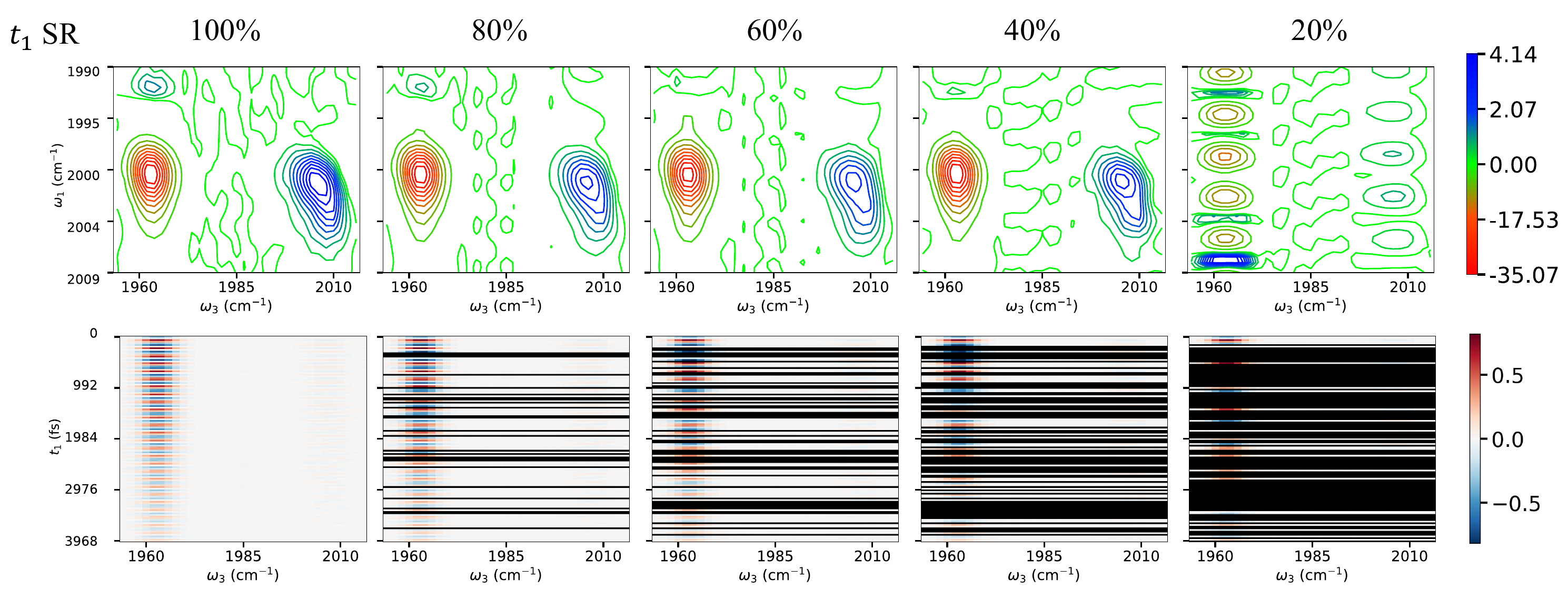}
    \caption{\textbf{Reconstruction performance at different $t_1$ sampling rates.} 
    We display the reconstructed 2DIR spectrum under different sampling rates and associated temporal data. SR: sampling rate.}
    \label{fig:drop_t1}
    \vspace{-3mm}
\end{figure}

\begin{figure}[t]
    \centering
    \includegraphics[width=\textwidth]{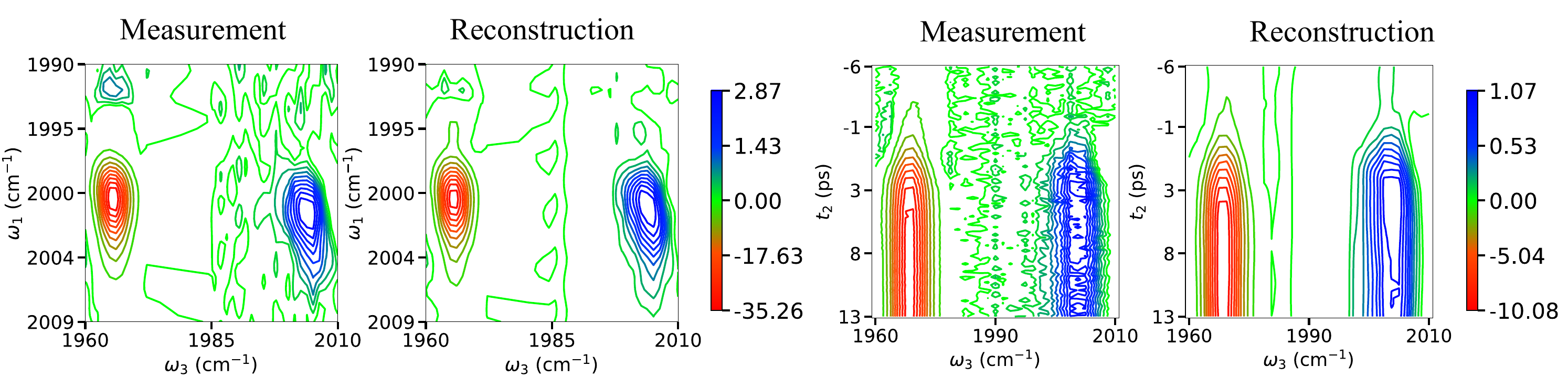}
    \caption{\textbf{Reconstruction 2DIR Spectrum under low accumulation and sampling rates.} 
    With $r=5$ and undersampled $t_1, t_2$ measurement, we display an unsampled $t_2$ slice of 2DIR spectrum and a slice of $(t_2, \omega_3)$ projection.}
    \label{fig:drop_t1t2}
    \vspace{-3mm}
\end{figure}

\begin{table}[t]
\centering
\caption{RFFT reconstruction error under different $t_1$ sampling rates.}
\label{tab:rfft_drop_mse_seed7}
\begin{tabular}{rcc}
\toprule
$t_1$ Sampling Rate & Spectrum MSE & Unsampled Temporal MSE \\
\midrule
80\% & \textbf{1.333e-02} & \textbf{1.583e-04} \\
60\% & 1.663e-02 & 1.824e-04 \\
40\% & 3.369e-02 & 3.825e-04 \\
20\% & 2.503e-01 & 2.429e-03 \\
\bottomrule
\end{tabular}
\end{table}

\subsubsection{Undersampled Coherent Temporal Data}
\label{ssec: subt1}

When undersampling the $t_1$ domain, we impose a sampling restriction that necessitates the inclusion of the first five and the final $t_1$ indices. The remaining indices in the undersampled set are then selected via random sampling from the intervening positions. Fig. \ref{fig:t1_results} evaluates signal reconstruction performance under a reduced $t_1$ sampling regime, illustrating (a) the 2DIR spectrum and (b) temporal measurements for the fully sampled ground truth, a 60\% sampling rate, and our proposed reconstruction. While lower sampling rates degrade spectral intensity due to truncated oscillations, our algorithm effectively recovers both the oscillation dynamics and the underlying spectral structures. This fidelity is further evidenced in Fig. \ref{fig:t1_results}(c), where the reconstructed continuous function closely aligns with the ground truth, accurately estimating values at unsampled points. To ensure a rigorous comparison, sinc interpolation was employed to recover the hidden continuous representation of the ground truth samples.

In Fig.~\ref{fig:t1_cmp}, we benchmark our proposed framework against two baseline approaches: (a) Fourier-feature encoding (FFE)~\cite{tancik2020fourier}, an MLP-based technique that takes raw coordinates and their high-frequency sinusoidal mappings as inputs to resolve rapid signal variations, and (b) GIRAF~\cite{bhattacharya2017accelerating}, an existing compressed sensing technique with a low-rank constraint developed for 2DIR. Although both baseline methods manage to recover the observed training points, they yield erroneous interpolation and hence fail to generalize to unobserved regions. Specifically, the FFE exhibits pronounced interpolation artifacts due to a lack of physical constraints. On the other hand, because GIRAF infers unobserved values through a local patch low-rank prior, the oscillatory structures in the time domain are easily oversmoothed under high missing rates. Furthermore, while GIRAF can recover 2DIR data from low $t_1$ sampling rates, it is unable to address undersampling in other dimensions, such as $t_2$. In contrast, our proposed RFFT-based model is capable of recovering 2DIR data when multiple dimensions ($t_1,t_2$) are undersampled. It accurately reconstructs dominant vibrational peaks while effectively suppressing aliasing artifacts induced by sparse sampling. Quantitatively, as summarized in Table~\ref{tab:drop50_avg_ang2_to_ang02}, this approach achieves the highest recovery accuracy, yielding lower errors in both unsampled regions and frequency-domain metrics.

We also investigated the impact of varying the subsampling rates along the $t_1$ axis, with the results summarized in Table \ref{tab:rfft_drop_mse_seed7} and Figure \ref{fig:drop_t1}. As observed, peak distortion becomes significant when 151 out of 251 samples are removed (40\%). In contrast to the slowly varying temporal data along the $t_2$ axis, our algorithm demonstrates a lower tolerance for sampling reduction in $t_1$. This discrepancy arises because the original sampling rate in $t_1$ is closer to the Nyquist limit; consequently, the loss of samples becomes critical for accurately recovering the high-frequency oscillating and decaying components of the temporal signal. Conversely, the recovery of slowly varying signals is inherently less sensitive to high subsampling rates, as the underlying dynamics are captured by fewer data points.

\subsubsection{Jointly Low Accumulation and Undersampled}

We evaluate the joint effects of a low signal-averaging count and undersampling across both the $t_1$ and $t_2$ dimensions on 2DIR signal recovery. Figure~\ref{fig:drop_t1t2} presents the reconstruction results using a $t_2$ sampling interval of 4, a $t_1$ sampling rate of 60\%, and a signal-averaging count of $r=5$, yielding 32-fold faster data collection. As observed in the omitted $t_2$ slices within the $(\omega_1, \omega_3)$ domain, our method successfully reconstructs a smooth yet structurally accurate 2DIR spectrum despite the elevated noise and sparse $t_1$ sampling. Furthermore, the underlying $t_2$ dynamics are faithfully recovered in this highly constrained regime. These results validate the robustness of our INR model against simultaneous noise and data sparsity in the measurement.


\subsection{Sampling Restrictions}
\begin{figure}[t]
    \centering
    \includegraphics[width=\textwidth]{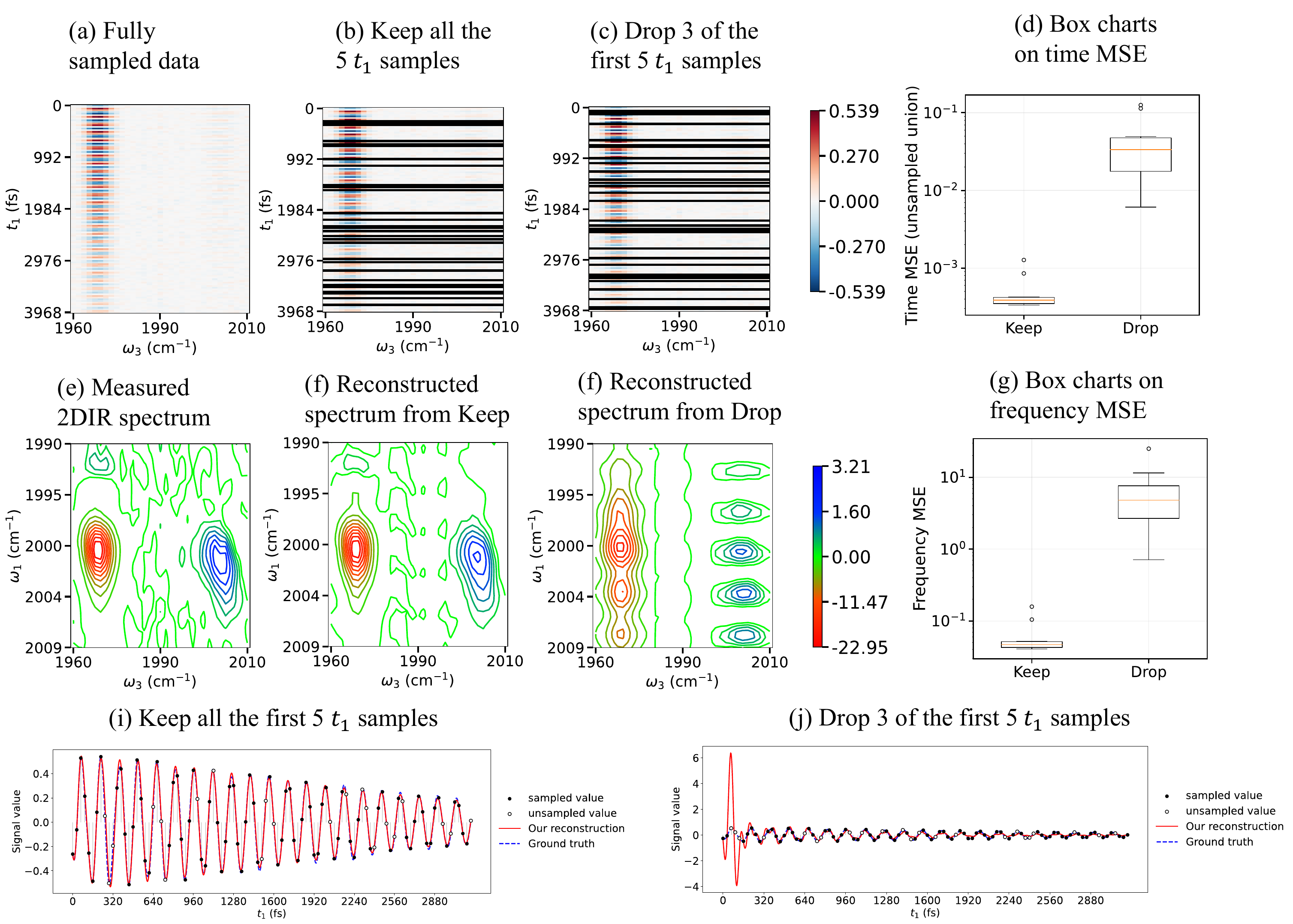}
    \caption{\textbf{Comparison of 2DIR Spectrum Reconstruction Under Differing Sampling Strategies.} 
    (a)-(c) Fully sampled temporal data and representative sampling masks for the two evaluated strategies. (d) Temporal reconstruction statistics across 10 independent trials. (e)-(g) Corresponding reconstructed 2DIR spectra for the examples shown in (a)-(c). (h) Summary of spectral reconstruction performance. (i)-(j) 1D illustrations of the reconstruction performance under the (i) Keep and (j) Drop strategies. Omitting the early $t_1$ samples leaves the reconstruction underconstrained, leading the model to hallucinate spurious initial oscillations.}
    \label{fig:sampling}
    \vspace{-3mm}
\end{figure}

In Fig.~\ref{fig:sampling}, we compare 2DIR spectrum reconstruction results from two distinct sampling strategies with identical total sample counts: one adhering to the early-time sampling restriction by retaining the first five $t_1$ samples (Keep), and one violating it by dropping three of these initial samples (Drop). We evaluate each strategy across 10 independent trials for robust statistics. Figures~\ref{fig:sampling}(a)-(c) display the fully sampled data alongside representative sampling masks, while Fig.~\ref{fig:sampling}(d) summarizes the temporal reconstruction results, demonstrating that the Keep strategy consistently outperforms the Drop strategy. The corresponding 2DIR spectra and spectral reconstruction performance are presented in Fig.~\ref{fig:sampling}(e)-(g) and Fig.~\ref{fig:sampling}(h), respectively. Notably, dropping the early $t_1$ samples introduces severe aliasing artifacts into the reconstructed spectra.

This performance gap arises from the interplay between the rapidly decaying nature of the temporal signal and the inductive biases of the INR. In the time domain of 2DIR spectroscopy, the earliest samples capture the highest signal amplitude and establish the absolute phase of the underlying oscillations \cite{zanni2020hyperspectral}. Drawing a parallel to nuclear magnetic resonance spectroscopy, omitting these initial high-amplitude samples severely underconstrains the signal, fundamentally introducing phase ambiguity into the spectral reconstruction \cite{ravera2021phase}. Algorithmically, INRs inherit the characteristic spectral bias of standard neural networks, prioritizing the learning of dominant, low-frequency structures such as the overall signal envelope \cite{rahaman2019spectral}. Because this initial decay defines the macroscopic signal envelope, the INR relies on these high-energy samples as its primary structural anchors; without them, the reconstruction problem becomes underspecified \cite{yuce2022structured}. When these boundary anchors are captured (Fig.~\ref{fig:sampling}(i)), the INR is sufficiently constrained to reliably establish the global phase. Conversely, without the precise alignment of these initial oscillations, the network must extrapolate the highest-amplitude onset. While Fourier-feature MLPs excel at interpolation \cite{tancik2020fourier}, they struggle to extrapolate; their harmonic basis functions often diverge unpredictably outside the training domain \cite{fesser2023understanding}. Furthermore, because the time and frequency domains are globally coupled, compensatory errors at this extrapolated boundary inevitably corrupt the entire sequence, manifesting as widespread spectral leakage and phase misalignment \cite{gottlieb1997gibbs}. Consequently, as shown in Fig.~\ref{fig:sampling}(j), the network overfits the low-amplitude tail while hallucinating unphysical variations at the poorly constrained boundary.

\subsection{Adaptive Sampling Results}
\label{ssec: adaptive}
\begin{figure}[H]
     \centering
     \includegraphics[width=\textwidth]{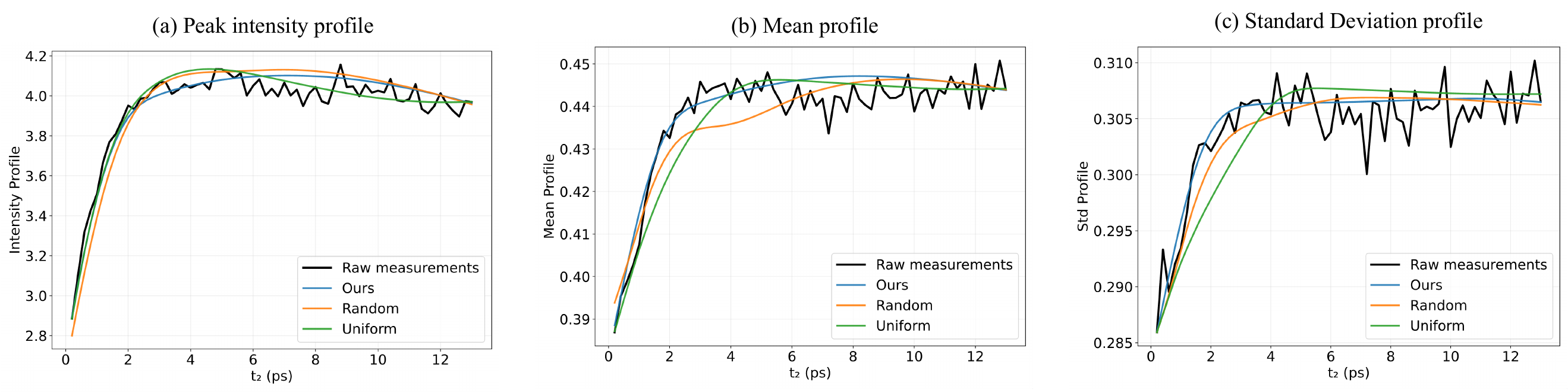}
     \caption{\textbf{Reconstruction results from different sampling strategies.} 
     We overlay (a) peak-intensity, (b) mean, and (c) standard-deviation profiles derived from spectra reconstructed using different sampling strategies.}
     \label{fig:adaptive_result}
     \vspace{-3mm}
\end{figure}

We compare multiple sampling strategies under a consistent experimental setup. Specifically, we first include the interval endpoints and then apply three strategies: the proposed loss-driven sampling, uniform sampling, and random sampling \cite{humston2017compressively, bhattacharya2017accelerating, humston2019optimized}. As summarized in Fig.~\ref{fig:adaptive_result} and Table~\ref{tab:active_clean}, the proposed loss-aware sampling achieves the best alignment to the intensity, mean, and standard deviation profiles from the measurement, as opposed to the other two strategies. Notably, all strategies utilize only 1/16 of the maximum sampling budget. As the sampling density increases beyond this regime, the performance gap between methods diminishes due to the generalization capability of the proposed neural representation framework.

\begin{table}[H]
\centering
\caption{Quantitative comparison of different adaptive sampling strategies.}
\label{tab:active_clean}
\begin{tabular}{lrrr}
\toprule
Method & Intensity MSE & Mean profile MSE & Std profile MSE \\
\midrule
Ours    & \textbf{6.190e-02} & \textbf{2.710e-06} & \textbf{2.320e-07} \\
Random  & 7.010e-02 & 4.950e-06 & 4.870e-07 \\
Uniform & 3.430e-01 & 4.870e-06 & 3.590e-07 \\
\bottomrule
\end{tabular}
\vspace{-3mm}
\end{table}

\subsection{Statistical-Moment Constraint}
\begin{figure}[t]
    \centering
    \includegraphics[width=\textwidth]{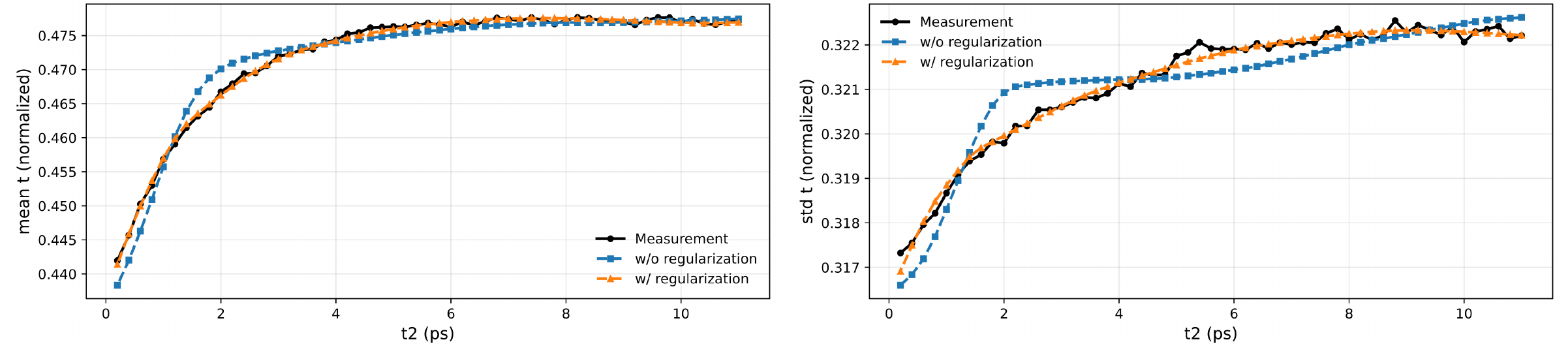}
\caption{\textbf{Effect of statistical moment-matching.} 
Spectral recovery results obtained with (orange) and without (blue) 
moment-matching constraints for (a) the mean and (b) the standard deviation 
profiles.}
    \label{fig:ablation}
    \vspace{-3mm}
\end{figure}

To validate the impact of the statistical-moment constraint detailed in Sec.~\ref{subsec:mlp}, we performed an ablation study by excluding the statistical and physical constraints from the optimization objective. As illustrated in Fig.~\ref{fig:ablation}, relying solely on $\mathcal{L}_{\mathrm{MSE}}$ results in significant deviations in the recovered spatial evolution profiles ($\mu_M$ and $\sigma_M$), even when the spectral intensities appear visually consistent with the ground truth. This discrepancy arises because $\mu_M$ and $\sigma_M$ are derived moments (Eq.~\ref{eq:mean_std}); unlike pixel-wise intensity, they are global descriptors highly sensitive to the probability distribution shape and slight spectral artifacts.

\subsection{Limitations and Possible Extensions}

The proposed method currently encounters scalability limitations when applied to extended 4DIR data acquisition ranges. Specifically, the efficacy of our normalization scheme diminishes when spectral lobes occupy only a minor fraction of the $(\omega_1, \omega_3)$ plane, as the spectral energy becomes compressed within a narrow interval of the normalized $[0,1]$ range. Potential solutions to this sparsity issue include adaptive coordinate networks, multi-resolution encodings~\cite{martel2021acorn, yu2021plenoctrees}, or importance-weighted training strategies that prioritize high-energy regions~\cite{raissi2019physics}. A further limitation involves extrapolation performance; although coordinate-based representations are well-suited for interpolation, they easily fail to extrapolate unsampled spectral coordinates, a common vulnerability in standard MLP architectures~\cite{chen2021learning, wei2024nermo}. Furthermore, in practical deployment, adaptive sampling requires frequent alternation between optimization and data acquisition, making the total elapsed time longer than theoretical estimates. We leave the optimization of this workflow as a direction for future research. Finally, while this study decouples oscillatory and non-oscillatory temporal variations, these dynamics may co-exist along the same dimension in complex systems; capturing such heterogeneous variations may necessitate further architectural refinements to the neural representation.

In addition to these methodological refinements, extending our model to other CMDS modalities offers significant potential for broader application. For instance, transient absorption microscopy~\cite{zhu2020transient, gross2023progress}, which also exhibits oscillatory and non-oscillatory population dynamics, could significantly benefit from this approach. Applying our INR-based framework to these slowly varying signals could decrease the number of samples, accelerating imaging while preserving the underlying dynamics.

\section{Conclusion}
The time required to collect high-quality 2DIR and other hypercube data remains a major challenge. To overcome this issue, we propose a physics-informed model that incorporates MLP-based spectral representation and a loss-aware adaptive sampling strategy. Experimental results show that the proposed model achieves high fidelity recovery using few data aggregation counts and sparse sampling along the population time axis. The probabilistic analysis demonstrates that the spectral dynamics are accurately preserved. In addition, by incorporating an IRFT-based optimization, our model manages to reconstruct rapid oscillations even with sparsely sampled signals along the coherence time axis. The incorporation of loss-driven active sampling further improves data efficiency. These advances highlight a path toward efficient, physics-aware acquisition of 2DIR, transient absorption microscopy and any hypercube data.

\section{Acknowledgement}
We acknowledge Mason Valentine for inspirations and technical support in this work. Wei Xiong and Harsh Bhakta are supported by Air Force Office of Scientific Research (FA9550-22-1-0317).







\end{document}